# CONVERGE – FUTURE IRS-INFRASTRUCTURES AS OPEN SERVICE NETWORKS

**Prof. Dr. Horst Wieker**

Hochschule für Technik und Wirtschaft des Saarlandes, University of Applied Sciences
Goebenstraße 40, 66117 Saarbrücken
(+49) 681 5867 195, wieker@htw-saarland.de

**Manuel Fünfrocken, Matthias Scholtes, Jonas Vogt, Niclas Wolniak**

Hochschule für Technik und Wirtschaft des Saarlandes, University of Applied Sciences
Altenkesseler Straße 17/D2, 66115 Saarbrücken

## ABSTRACT

In the past years there has been significant progress in Intelligent Transport Systems in the domains of traffic management, driver assistance and driver information. Within the German research project COmmunication Network VEhicle Road Global Extension (CONVERGE), a system architecture for the flexible interaction between different service providers and communication network operators will be designed and implemented, called Car2X Systems Network architecture. The Car2X Systems Network will establish a completely new open communication-, services-, and organization architecture that reflects state of the art communication technologies and IT security. Through well-defined access methods, service providers like traffic control centers or vehicle manufacturers can be integrated into the open and secure system network. The ultimate goal is the decentralized and dynamic coupling of all systems and actors across national and organizational borders. This paper is focusing on the requirements and the concept of a service-oriented architecture for publishing and finding services inside the system with intelligent information distribution based on geographical and organizational conditions.

## MOTIVATION

With Intelligent Transport Systems (ITS) at the edge of deployment, pioneering approaches to traffic management and vehicle safety issues are increasingly growing together. Still a holistic system architecture for flexible interaction between different service providers and communications network operators is missing in a decentralized, scalable structure. The aim of the German research project CONVERGE [1] is to close this gap.

The developed Car2X Systems Network (see Figure 1: CONVERGE Car2X Systems Network) is a dynamical extendable association for cooperative systems in ITS, comparable with the internet with open standards and interoperability, which picks up current innovations in the field of information and communication technology (ICT) and thus provides novel approaches for system- and software-design.

The stated goal of CONVERGE is the definition of an architecture and interface for a Car2X Systems Network. This network is supposed to be open, providing its integration interfaces to the international standardization process as well as scalable and flexible, to ensure a successful rollout. The entities of the network are distributed and provider-independent, allowing for a flexible role model in which several participants can fulfill a role, lowering financial barriers to participate. The system shall not be limited by national borders, enabling a trans-regional and international deployment. To account for high requirements and standards in the field of IT security, this is considered and actively included in the developments for each component as well as the overall system design. The innovative approach of a hybrid communicating network encompasses the communication between backend components and mobile ITS stations like vehicles via various access technologies. Currently this includes ETSI ITS G5, a wireless standard specialized for ITS, and cellular communication.

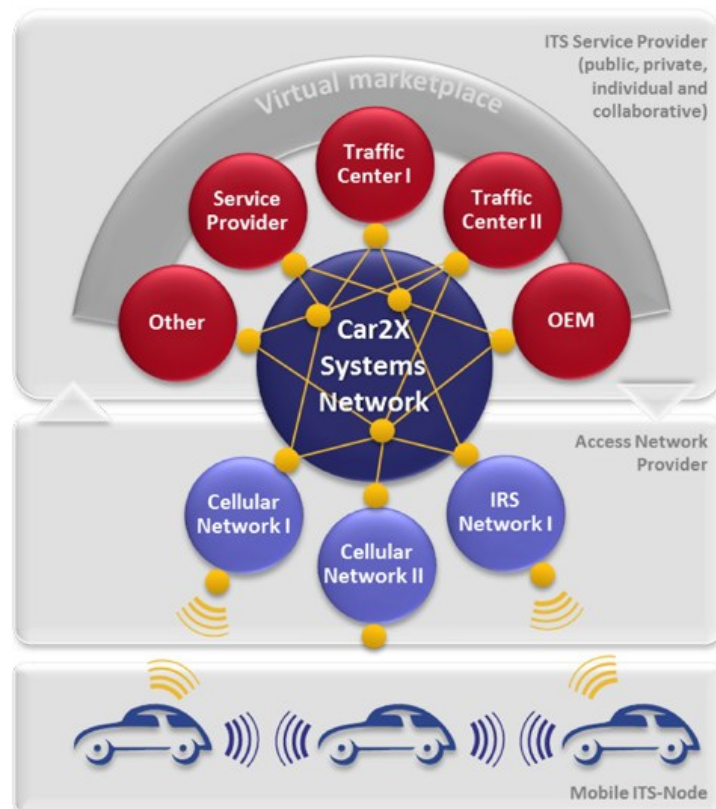

**Figure 1: CONVERGE Car2X Systems Network**

CONVERGE, as a service-oriented architecture, should provide an easy way to publish and lookup services for participants in the Car2X Systems Network. They are provided by Service Providers (SP) and utilized by Service Customers (SC). Services themselves can be distinguished into global services (GS) and local services (LS). While GS are supposed to be relevant for every participant independent of its actual position, LS may only be relevant for participants within a certain geographical area. Examples for such LS are parking area information, petrol prices, detour notifications and road works information. For example, a parking lot could offer a service, providing information about its capacity and occupancy via an attached ITS Roadside Station (IRS). This information is primarily relevant within the communication range of the respective IRS. With increasing distance, the significance of the service may decrease.

In the following, the concept of a Service Directory (SD) and its beneficial distribution related to a geographical area will be explained. This paper focusses on concepts regarding IRS networks as communication networks exclusively. An IRS is a permanently or semi-permanently unit installed on the roadside, which is able to communicate with passing vehicles over ETSI ITS G5 [2], forming an infrastructure for C2X communication. Within the project CONVERGE, all IRS are based on the standardized architecture of the Roadside ITS Substations according to ETSI EN 302 665 [3].

**CONCEPT**

The SD is a service provided by a Car2X Systems Network internal functionality. It is used to publish new services and provides a database to find services in the Car2X Systems Network. The SD may be realized in a decentralized structure (see Figure 2: (Distributed) Service Directory Architecture) in which most nodes hold only a subset of service entries and collaborative searches are applied over a couple of nodes. Moreover, feasible algorithms for service-database distribution and synchronization as well as for efficient search have to be applied. This decentralized organization allows for a provider open organization and hierarchical configuration of the SD. A requirement from the CONVERGE project foresees, that these mechanisms, especially synchronization, provide a high quality in terms of topicality, speed and reliability.

A participant, who offers a service, acts as SP. To publish their services, SP contact the SD to register the respective service as well as related meta-information in a searchable database. The SD serves as a database that allows filtered searches to look up available services, respectively their definition, containing data types and formal access description (e.g. WSDL [4]). The SD database stores substantial information about the published services and acts as a broker between SC and SP. Additionally SC can use the SD to lookup a service with desired functionality. A service itself is a combination of both data processing and data exchange for a defined purpose.

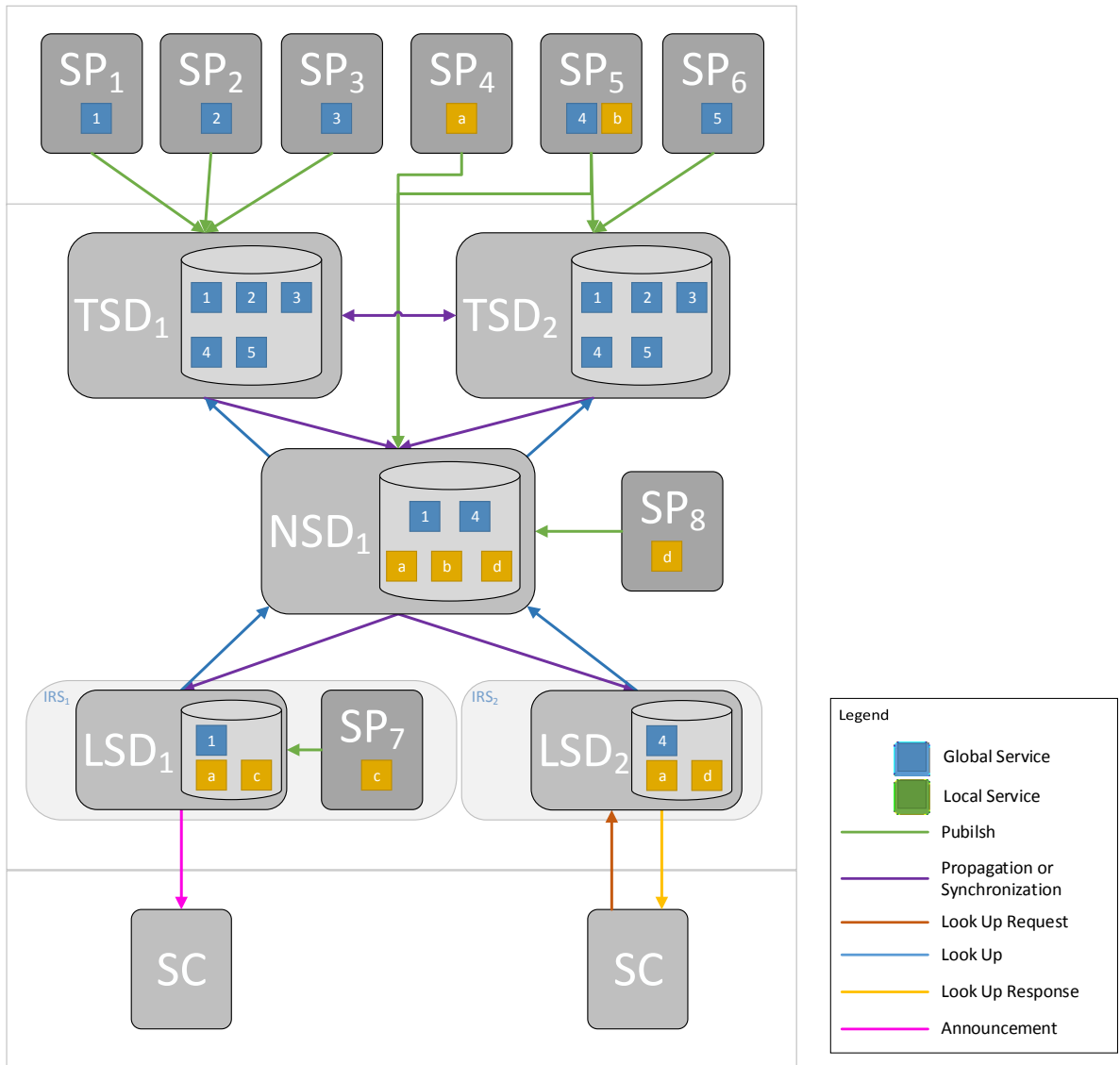

**Figure 2: (Distributed) Service Directory Architecture**

**Hierarchy**

As mentioned in the motivation, some service information are only relevant for geographical areas. Therefore, we consider the approach of a distributed SD, related to its geographical relevance. The distribution follows a layered hierarchy (see Figure 2: (Distributed) Service Directory Architecture):

- Top Level Service Directory (TSD) Layer:
  Global Service Directories are placed at the Top Level (Top Level Service Directory TSD). Within that level, which is placed in the backend of the Car2X Systems Network, we assume that every SD is completely synchronized with the other SD. Hence every TSD encompasses the full set of GS. Regarding the previously mentioned example of the parking lot, the service at this layer could be an overregional SP, offering parking lot coordination services to SC.

- Network Service Directory (NSD) Layer:
  Following the hierarchy according to our focus on IRS networks, the next layer consists of a set of Service Directories containing a subset of GS. They are placed in the backend of an IRS network. Therefor they are called Network Service Directories (NSD). From a SD point of view, they divide a physical IRS network into several logical networks. It is possible for the NSD level to consist of various parallel NSD entities creating the same amount of logical subnetworks. Nevertheless, a full hierarchy of NSD is also possible in order to structure a network even more. For example, a physical network covering a federal state could be logically divided into four networks for the regions north, south, west and east using four NSD. Below may be a number of subnetworks for individual municipalities and cities. With respect to the parking lot example, at this position the service is registered to provide access not only to the respective LSD at the parking lot location but also to nearby areas, where SC can use the service to feed route planning services with end-of-route parking information.

- Local Service Directory (LSD) Layer:
  At the bottom level of the hierarchy resides the Local Service Directories (LSD). A LSD entity is placed within an IRS and belongs to a logical subnetwork created by the overlaying NSD. Inside an IRS, an SP or a part of an SP can be located. This SP publishes it service directly inside the LSD located on its IRS. In this case the provided service is classified as LS. The exemplary parking lot service is restricted to the communication area of the respective IRS. The service information is not distributed to other SD, not even to the LSD of neighboring IRS or the parent NSD.

**Registration/Publication**

The registration of GS takes place at the TSD exclusively. The registration of LS depends on several factors and can be processed at the NSD or LSD. As a general rule, a service may only be propagated to SDs below its originating SD, so a service registered on an NSD is only available to the LSD below it.

**Propagation**

The propagation of services within the SD hierarchy is always unidirectional from upper layers to lower layers. For GS this is implicitly true due to the TSD being the only point of registration for GS. LS, which are registered at the NSD, can be propagated towards connected LSD. LS, which are registered at the LSD, are therefore restricted to their IRS and cannot be propagated to other SD. To make LS available to other LSD, they have to be registered at a NSD above the respective LSD.

To distribute service information and make them available to the SC we propose two concepts to propagate service information between the layers.

- Persistent preemptive propagation (PPP):
  The service information is pushed from source to sink (TSD to NSD or NSD to LSD) without being queried initially. The information are stored persistently or with a high caching time (see Figure 3: Persistent preemptive propagation (PPP)).

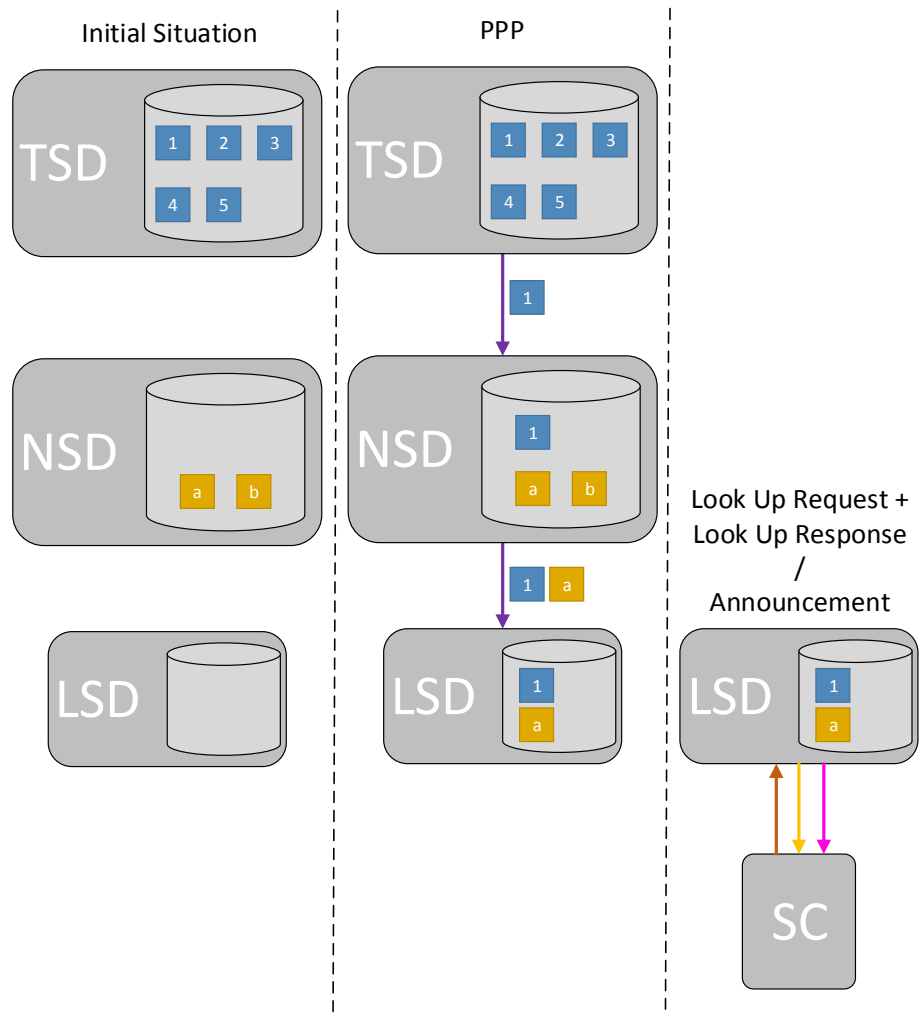

**Figure 3: Persistent preemptive propagation (PPP)**

- Volatile reactive propagation (VRP):
  The service information is requested by the sink (LSD to NSD or NSD to TSD), reacting to a request, initiated by a SC. The source propagates the requested information afterwards to the sink if available comparable to the process in PPP. The information is cached, with the caching time being increased or refreshed by further incoming requests (see Figure 4: Volatile reactive propagation (VRP)).

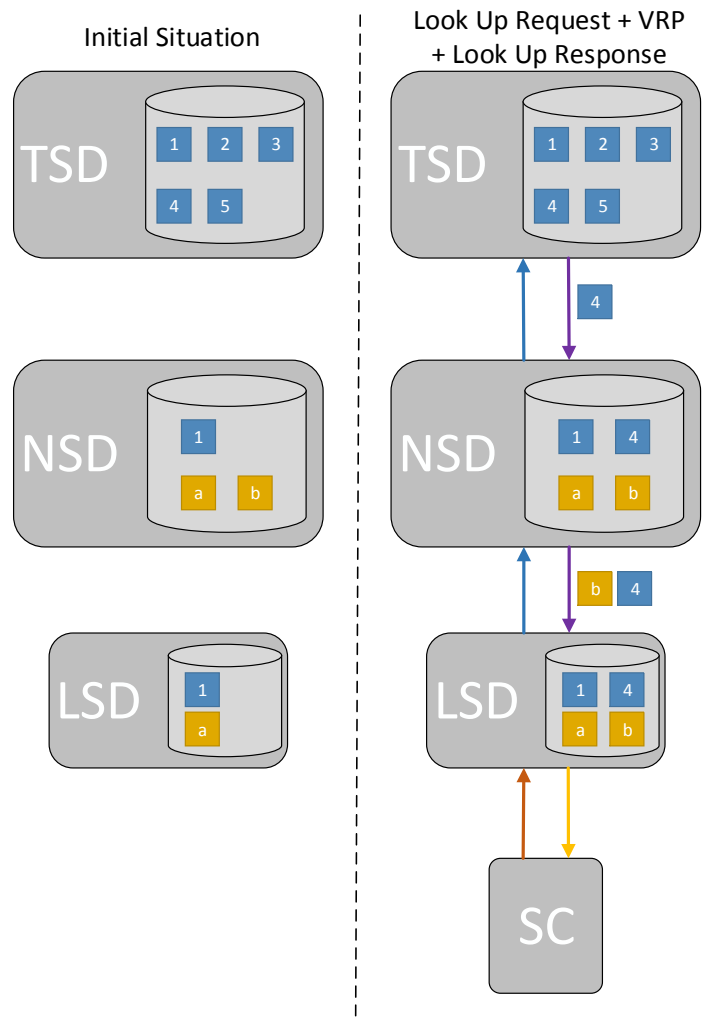

**Figure 4: Volatile reactive propagation (VRP)**

These concepts are applied to specific services. An entity of the layer may have to be able to support both in order to handle services, which require PPP as well as VRP.

**Synchronization**

The synchronization of service information follows the propagation of the respective service and is therefore processed between SD of different layers. A propagation source synchronizes service information towards a propagation sink. An exception for this behavior is the synchronization between TSD. In order to provide redundant data via all TSD, service information are synchronized and not propagated between TSD. The synchronization includes the deletion of service information.

**Announcement/Discovery**

If services are announced or discovered, this is always located at the bottom layer, between LSD and SC. A SC will never directly communicate with a NSD or TSD.

## VISION

With the previous shown concept, it will be possible to provide access to geographical aligned services utilizing subtrees of the SD hierarchy. The entities are physically distributed among one or more IRS located in the respective area as well as the backend of the network. This concept leads to a convergence of information to its associated geographical area, hence higher layer SD are less likely to be queried. This results in reduced traffic and processing load for the communication towards the TSD.

The concept will offer new possibilities to distribute geo-referenced service information. An LSD may identify the relevance of service information for its geographical area by analyzing utilization. The SD hierarchy may afterwards decide to change the propagation strategy for that service. A VRP propagated service that is highly used may be better suitable for a PPP propagation and vice versa.

In a later step, one could use statistical analysis of the local SD to identify services, which only have local relevance and a huge amount of consumers and decide to put instances of those services on IRS or move the SP closer to those IRS, thus reducing the overall network load and global resource usage of the service. This steps could even be performed automatically.

The concept of distributed SD, georeferenced services and their alignment to local SD descends from the conception work in the CONVERGE project. The content of this paper summarizes our approaches to address upcoming challenges developing a scalable solution for service distribution with regards to a later rollout scenario. Therefore the presented concepts will be incorporated creating the solutions in CONVERGE as well as other test fields like ITeM (ITS Test field Merzig) [5].

## ACKNOWLEDGEMENT

This work was funded within the project CONVERGE by the German Federal Ministries of Education and Research as well as Economic Affairs and Energy. The results presented in this paper were developed jointly by the CONVERGE project partners. The CONVERGE consortium consists of: Adam Opel AG, Federal Highway Research Institute (Bundesanstalt für Straßenwesen), BMW Research and Technology GmbH, Robert Bosch GmbH, Ericsson GmbH, Fraunhofer Institute for Open Communication Systems (FOKUS), Fraunhofer Research Institution for Applied and Integrated Security (AISEC), Hessen Mobil Road and Traffic Management, PTV Planung Transport Verkehr AG, Vodafone GmbH, Volkswagen AG, Road administration of the city of Frankfurt/Main, Federal Network Agency (Bundesnetzagentur) and Hochschule für Technik und Wirtschaft des Saarlandes - University of Applied Sciences (project coordinator).